\documentclass[twocolumn,showpacs,preprintnumbers,amsmath,amssymb,prb]{revtex4}
\usepackage{graphicx}% Include figure files
\usepackage{dcolumn}% Align table columns on decimal point
\usepackage{bm}% bold math

\begin{document}

%\preprint{APS/123-QED}

\title{K$_2$Cr$_8$O$_{16}$ predicted as a half-metallic ferromagnet:\\ 
Scenario for a metal-insulator transition}

\author{M. Sakamaki}
\author{T. Konishi}
\affiliation{Graduate School of Advanced Integration Science, 
Chiba University, Chiba 263-8522, Japan}
\author{Y. Ohta}%\email{ohta@science.s.chiba-u.ac.jp}
\affiliation{Department of Physics, Chiba University, 
Chiba 263-8522, Japan}

\date{June 4, 2009}% It is always \today, today,
             %  but any date may be explicitly specified

\begin{abstract}
Based on the first-principles electronic structure calculations, 
we predict that a chromium oxide K$_2$Cr$_8$O$_{16}$ of hollandite 
type should be a half-metallic ferromagnet where the Fermi level 
crosses only the majority-spin band, whereas the minority-spin 
band has a semiconducting gap.  We show that the double-exchange 
mechanism is responsible for the observed saturated ferromagnetism.  
We discuss possible scenarios of the metal-insulator transition 
observed at low temperature and we argue that the formation of 
the incommensurate, long-wavelength density wave of spinless 
fermions caused by the Fermi-surface nesting may be the origin 
of the opening of the charge gap.  
\end{abstract}

%\keywords{Suggested keywords}

\maketitle

%%%%%%%%%%%%%%%%%%%%%%%%
\section{Introduction}
%%%%%%%%%%%%%%%%%%%%%%%%

Half-metallic ferromagnets \cite{katsnelson} offer a unique 
opportunity for studying the electronic states of strongly 
correlated electron systems.  Here, only the majority-spin 
electrons form the Fermi surface with a gapped minority-spin 
band \cite{note} and can couple with excitations of the spin 
(and in some cases orbital) degrees of freedom of the system.  
The situation therefore should attract much interest, in 
particular, when the relevant electrons are strongly 
correlated, leading the system to double-exchange 
ferromagnetism \cite{zener} and metal-insulator 
transition (MIT).  
In this paper, we will show that a chromium oxide 
K$_2$Cr$_8$O$_{16}$ with the hollandite-type crystal 
structure \cite{K2Cr8O16,Rb2Cr8O16} belongs to this class of 
materials and can provide a good opportunity for further 
development of the physics of strong electron correlations.  

The crystal structure of K$_2$Cr$_8$O$_{16}$ (see Fig.~1) 
belongs to a group of hollandite-type phases where 
one-dimensional (1D) double strings of edge-shared CrO$_6$ 
octahedra forms a Cr$_{8}$O$_{16}$ framework of a tunnel 
structure, wherein K ions reside.\cite{tamada}  
Cr ions are in the mixed-valent state of 
Cr$^{4+}$ ($d^2$) : Cr$^{3+}$ ($d^3$) $=3:1$, and hence 
with 2.25 electrons per Cr ion.  
It has recently been reported \cite{hasegawa} that the phase 
transition from the Pauli-paramagnetic metal to ferromagnetic 
metal occurs at $T_c\simeq 180$ K by lowering temperatures, 
where the ferromagnetic state has a full spin-polarization 
of 18 $\mu_{\rm B}$ per formula unit (f.u.) at low 
temperatures, which is a realization of saturated ferromagnetism.  

In addition to this phase transition, it has also been 
reported \cite{hasegawa} that another phase transition occurs 
from the ferromagnetic metal to ferromagnetic insulator at 
$T_{\rm MI}\simeq 80$ K, suggesting that the charge gap opens 
below $T_{\rm MI}$.  Surprisingly, the spin polarization is 
hardly affected by this MIT and no structural distortions 
associated with this MIT have been observed so 
far.\cite{hasegawa}  
The mechanism of the MIT of this material has therefore been 
a great puzzle.  

\begin{figure}[tbh]
\begin{center}
\resizebox{8.0cm}{!}{\includegraphics{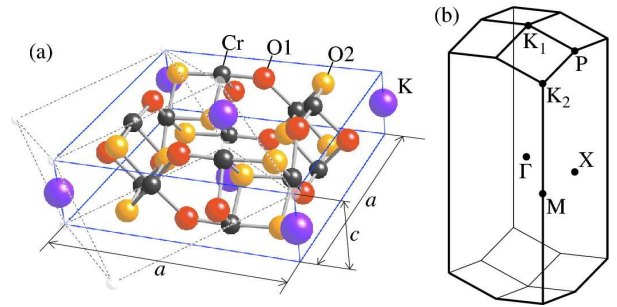}}\\
\caption{(Color online) Schematic representation of 
(a) the unit cell of the body-centered tetragonal lattice 
(solid lines) and (b) Brillouin zone of K$_2$Cr$_8$O$_{16}$.  
In (a), the primitive unit cell is also shown in the 
thin dotted lines.  In (b), the symbols represent 
$\Gamma(0,0,0)$, M$(2\pi/a,0,0)$, X$(\pi/a,\pi/a,0)$, 
P$(\pi/a,\pi/a,\pi/c)$, K$_1$$(0,0,\pi(1/c+c/a^2))$, 
and K$_2$$(2\pi/a,0,\pi(1/c-c/a^2))$, where K$_1$ and 
K$_2$ are equivalent.}
\label{fig.1}
\end{center}
\end{figure}

In this paper, we perform the first-principles electronic 
structure calculations based on the generalized gradient 
approximation (GGA) in the density-functional theory (DFT) 
and we predict that the materials $A_2$Cr$_8$O$_{16}$ (A = K 
and Rb) belong to a new class of half-metallic ferromagnets, 
i.e., the majority-spin electrons are metallic, whereas the 
minority-spin electrons are semiconducting with a band gap.  
We also show from the GGA and GGA+$U$ calculations that 
the double-exchange mechanism is responsible for the observed 
saturated ferromagnetism.  
We then discuss possible mechanisms of the MIT of this material 
and argue that the formation of an incommensurate, 
long-wavelength spin and charge density wave (DW) due to 
Fermi-surface nesting may be the origin of the MIT of this 
material.  
The opening of the gap in the majority-spin band should, 
however, be detrimental to the double-exchange ferromagnetism, 
for which we consider possible reconciliations.  
We hope that our results will encourage further experimental 
studies of this intriguing material.  

This paper is organized as follows:  
In Sec.~II, we show our method of calculations.  
In Sec.~III, we present our results of calculations and 
discuss the origin of ferromagnetism and possible mechanisms 
of the MIT of K$_2$Cr$_8$O$_{16}$.  
Summary of our work and prospects for future studies are 
given in Sec.~IV.  

%%%%%%%%%%%%%%%%%%%%%%%%%%%%%%%%%
\section{Method of calculation}
%%%%%%%%%%%%%%%%%%%%%%%%%%%%%%%%%

The electronic structure calculation in GGA is carried 
out by employing the computer code WIEN2k,\cite{wien2k} 
which is based on the full-potential linearized 
augmented-plane-wave (FLAPW) method.  We use the 
exchange-correlation potential of Ref.\cite{PBE96}.  
The spin polarization is allowed.  
The spin-orbit interaction is not taken into account.  
We use 1,221 ${\bm k}$ points in the irreducible part 
of the Brillouin zone in the self-consistent calculations.  
We use the plane-wave cutoff of 
$K_{\rm max}=4.24$ Bohr$^{-1}$.  
The GGA+$U$ calculation \cite{anisimov} is also made to 
see the effects of on-site electron correlation $U$ on 
the band structure.  
We assume the experimental crystal structure of 
K$_2$Cr$_8$O$_{16}$ observed at room temperature with 
the lattice constants of $a=9.7627$ and 
$c=2.9347$ \AA.\cite{tamada}  
The Bravais lattice is body-centered tetragonal and the 
primitive unit cell (u.c.) contains four Cr ions, one K 
ion, and eight O ions, i.e., KCr$_4$O$_8$.  
The local structure around Cr ions is shown in Fig.~2.   
We use the code XCrySDen\cite{kokalj} for graphical 
purposes.  

\begin{figure}[tbh]
\begin{center}
\resizebox{6.5cm}{!}{\includegraphics{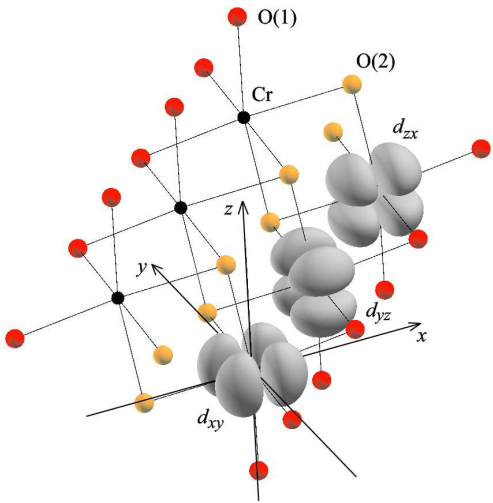}}\\
\caption{(Color online) Schematic representations of the 
local structure of K$_2$Cr$_8$O$_{16}$, together with the 
three $t_{2g}$ orbitals $d_{xy}$, $d_{yz}$, and $d_{zx}$ 
of Cr ions in the $xyz$-coordinate system.  There are two 
inequivalent O sites, O(1) and O(2).  All the Cr sites are 
equivalent.}
\label{fig2}
\end{center}
\end{figure}

%%%%%%%%%%%%%%%%%%%%%%%%%%%%%%%%%
\section{Results and discussion}
%%%%%%%%%%%%%%%%%%%%%%%%%%%%%%%%%

\subsection{Ground state}

The total ground-state energies obtained in GGA are 
$-10816.1918$ Ryd/u.c.~in the ferromagnetic state 
and $-10815.9597$ Ryd/u.c.~in the paramagnetic state, 
resulting in the energy gain of $3.16$ eV/u.c.~due to 
stabilization by the spin polarization, which is 
realized already in the GGA level.  
In the ground state, we have the full spin polarization 
with the magnetic moment of $9.000$ $\mu_{\rm B}$/u.c., 
which consists of the contributions from 
Cr ions, 2.043 $\mu_{\rm B}$/atom, 
O(1) ions, 0.014 $\mu_{\rm B}$/atom, and 
O(2) ions, $-0.082$ $\mu_{\rm B}$/atom, in the 
atomic spheres, and $1.098$ $\mu_{\rm B}$/u.c., 
in the interstitial region.  
Note that the O(2) ions inside the double strings of Cr 
ions have a negative spin polarization, whereas the O(1) 
ions connecting the double strings have a positive spin 
polarization.  

\subsection{Density of states}

The calculated density of states (DOS) is shown in Fig.~3 
in a wide energy range covering over the O $2p$ and 
Cr $3d$ bands.  We find three separate peaks in 
both the majority and minority spin bands: in the 
majority (minority) spin band, the O $2p$ weight is 
located mainly at $-7.5\lesssim\varepsilon\lesssim-1.8$ 
($-7.1\lesssim\varepsilon\lesssim-1.8$) eV, 
Cr $3d$ weight with the $t_{2g}$ symmetry at 
$-1.7\lesssim\varepsilon\lesssim 0.7$ 
($0.7\lesssim\varepsilon\lesssim 2.6$) eV, and 
Cr $3d$ weight with the $e_g$ symmetry at 
$1.7\lesssim\varepsilon\lesssim 4.0$ 
($2.9\lesssim\varepsilon\lesssim 5.1$) eV.  
The hybridization between the O $2p$ and Cr $3d$ 
bands is significantly large.  
The Fermi level is located at a deep valley of the 
$t_{2g}$ majority-spin band while it is located 
in the energy gap between the O $2p$ and Cr $3d$ 
$t_{2g}$ bands in the minority-spin band.  
Thus, the half metallicity of this material is evident 
in the calculated DOS.  

\begin{figure}[tbh]
\begin{center}
%\resizebox{7.0cm}{!}{\includegraphics{fig3a.eps}}\\
%\resizebox{7.0cm}{!}{\includegraphics{fig3b.eps}}\\
%\resizebox{7.0cm}{!}{\includegraphics{fig3c.eps}}\\
\resizebox{7.5cm}{!}{\includegraphics{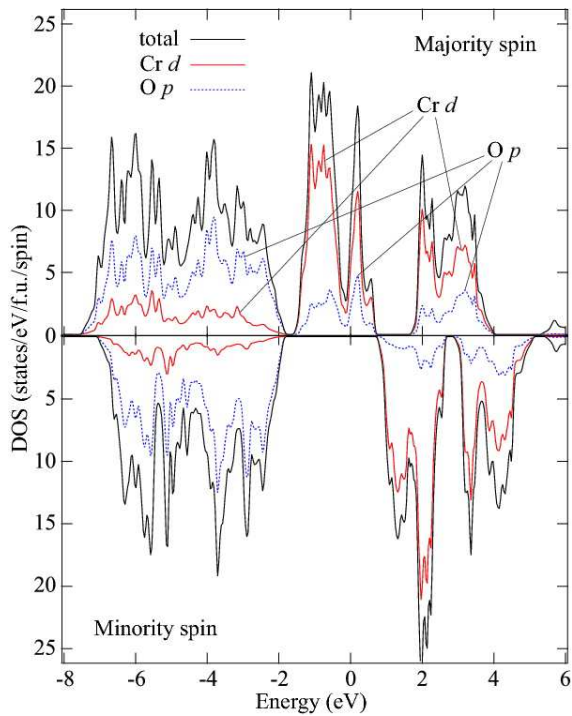}}\\
\caption{(Color online) Calculated result for the DOS 
(per spin per formula unit (f.u.)) of K$_2$Cr$_8$O$_{16}$ 
in a wide energy range for 
%(a) the paramagnetic state and 
%(b) majority-spin and 
%(c) minority-spin band in the ferromagnetic state.  
the ferromagnetic state.  
The Fermi level is set to be the origin of energy.}
\label{fig3}
\end{center}
\end{figure}

The calculated orbital-decomposed partial DOS, 
$\rho_\alpha(\varepsilon)$ ($\alpha=xy, yz, zx$), 
in the Cr $3d$ $t_{2g}$ region are shown in Fig.~4, 
where the two components are exactly degenerate, 
$\rho_{yz}(\varepsilon)=\rho_{zx}(\varepsilon)$, 
for both paramagnetic and ferromagnetic states.  
The three $t_{2g}$ orbitals are almost equally 
occupied by electrons in the paramagnetic state.  
In the ferromagnetic state, however, the $d_{xy}$ orbitals 
is almost fully occupied by electrons and therefore holes 
are only in the $d_{yz}$ and $d_{zx}$ orbitals.  
Also, the $d_{xy}$ component has a rather high peak-like 
structure at $\sim$0.7 eV below the Fermi level, 
indicating an essentially localized character of 
the $d_{xy}$ electrons.  The $d_{yz}$ and $d_{zx}$ 
components, on the other hand, have a rather wide band 
spreading over $-0.2\lesssim\varepsilon\lesssim 0.7$ eV 
around the Fermi level, indicating an itinerant character 
of the $d_{yz}$ and $d_{zx}$ electrons.  Here, the 
admixture of the $2p_z$ state of O(2) is significantly 
large: it occurs between the $2p_z$ orbitals of O(2) 
and the $d_{yz}$ and $d_{zx}$ orbitals of its neighboring 
Cr ions (see Fig.~2).  
These results suggest that the double-exchange mechanism 
\cite{zener} should work for the occurrence of 
ferromagnetism, as will be discussed further below.  

\begin{figure}[tbh]
\begin{center}
\resizebox{7.0cm}{!}{\includegraphics{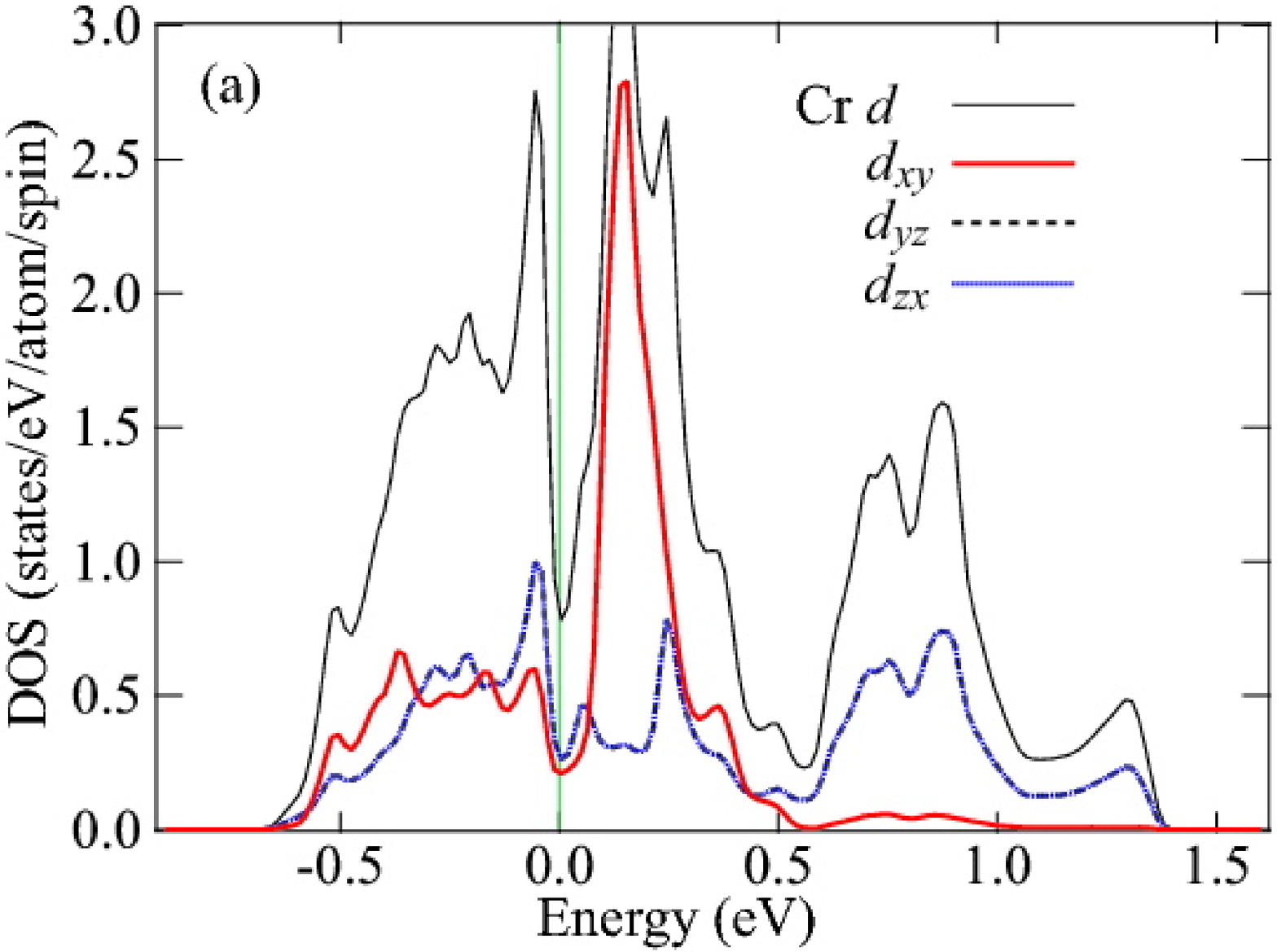}}\\
\resizebox{7.0cm}{!}{\includegraphics{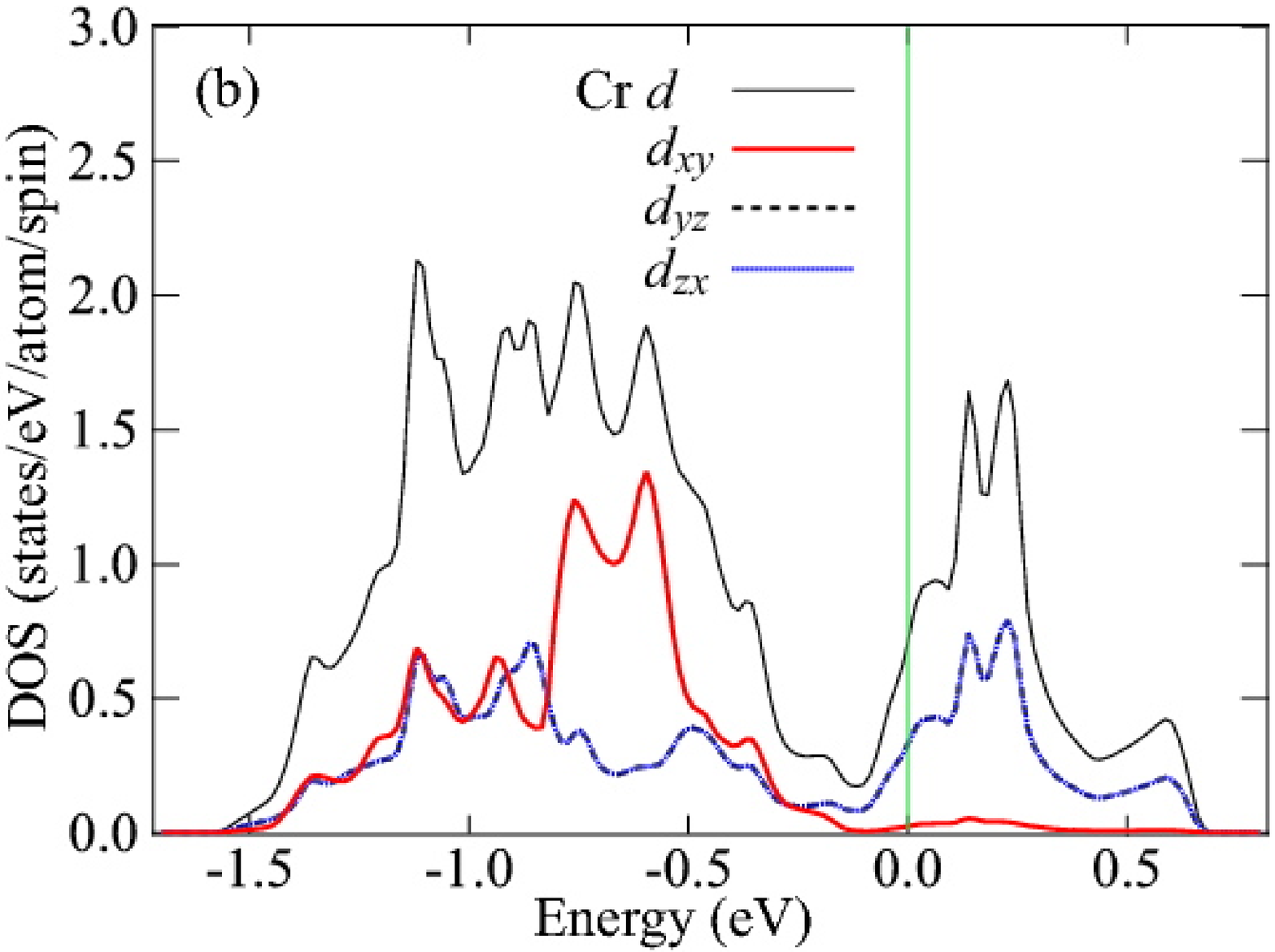}}\\
\caption{(Color online) Calculated orbital-decomposed 
DOS (per spin per formula unit (f.u.)) of K$_2$Cr$_8$O$_{16}$ 
near the Fermi level for the (a) paramagnetic state and 
(b) ferromagnetic state.  The Fermi level is indicated by 
the vertical line.  Contributions from the $d_{yz}$ and 
$d_{zx}$ orbitals (thin dashed and dotted lines) are 
exactly degenerate.}
\label{fig4}
\end{center}
\end{figure}

\subsection{Band dispersion}

The situation may be clarified further if one observes 
the calculated band dispersion near the Fermi level.  
The result is shown in Fig.~5.  
We find that a rather dispersionless narrow band of 
predominantly $d_{xy}$ character is located at $\sim$0.7 eV 
below the Fermi level, extending over a large region of 
the Brillouin zone.  On the other hand, the dispersive 
$t_{2g}$ bands of predominantly $d_{yz}$ and $d_{zx}$ 
character with strong admixture of the $2p_z$ state of 
O(2) are located at $-0.2\lesssim\varepsilon\lesssim 0.7$ 
eV and cross the Fermi level.  
Thus, we have the dualistic situation where the essentially 
localized $d_{xy}$ electrons at $\sim$0.7 eV below the 
Fermi level interact with the itinerant $d_{yz}$ and 
$d_{zx}$ electrons of the bandwidth comparable with the 
intraatomic exchange energy of $\sim$1 eV, whereby the 
Hund's rule coupling gives rise to the ferromagnetic 
spin polarization via the double-exchange 
mechanism.\cite{zener}

To support this further, we make the GGA+$U$ calculation 
for the present material (of which the results are not 
shown here).  We find that, as $U$ increases, the 
$d_{xy}$ band shifts further away from the Fermi level, 
leaving essentially no weight above the Fermi level, 
whereas the $d_{yz}$ and $d_{zx}$ bands with strong 
admixture of the O(2) $2p_z$ states are much less 
affected by the presence of $U$.  
These results are consistent with what is expected in 
the double-exchange mechanism of ferromagnetism.  
We point out that the situation is very similar to the 
case of CrO$_2$, which has so far been discussed in 
detail.\cite{schwarz,korotin,yamasaki} 

Another aspect noted in Fig.~5 is that we have the 
semi-metallic band structure in the sense that the number 
of electrons is equal to the number of holes near the Fermi 
level (also see Fig.~6 below); there are 12 bands for the 
$t_{2g}$ orbitals in the unit cell, which are fully 
spin-polarized and are occupied by 9 ``spinless fermions'', 
i.e., by 9 up-spin (or majority-spin) electrons.  
The integer filling of spinless fermions can give rise to 
either semiconducting or semimetallic band structure, 
of which the latter is realized in the present system.  
This result suggests that the MIT observed in 
K$_2$Cr$_8$O$_{16}$ can possibly be a 
semimetal-to-semiconductor transition, where the band 
gap opens between the lower 9 bands and upper 3 bands 
due to any unknown reasons.  To realize this, however, we 
need the symmetry reduction (due to lattice distortion) 
because the band degeneracy between the third and fourth 
bands (counted from the top of the 12 $t_{2g}$ bands) at 
the P point of the Brillouin zone cannot be lifted without 
breaking the 4-fold rotational symmetry around the $c$ axis 
of the tetragonal lattice.  Without the lattice distortion, 
we could have a zero-gap semiconductor, but the experimental 
data on the temperature dependence of the electrical 
resistivity \cite{hasegawa} suggest that this possibility 
should be ruled out.  Also, because no indications of this 
type of lattice distortions at the MIT have been observed 
so far,\cite{hasegawa} the semimetal-to-semiconductor 
transition seems to be unrealistic in K$_2$Cr$_8$O$_{16}$.  

\begin{figure}[tbh]
\begin{center}
\resizebox{8.0cm}{!}{\includegraphics{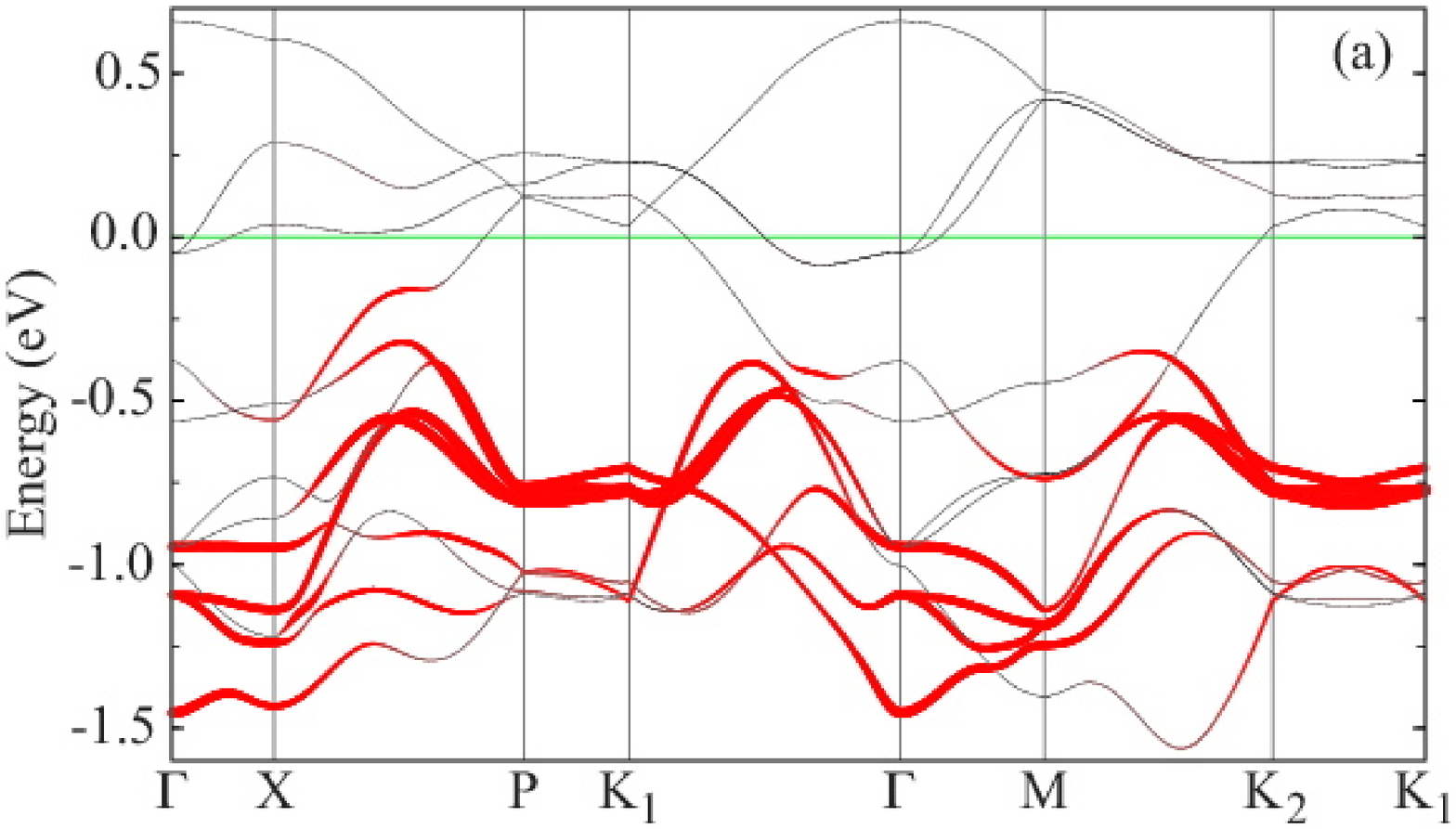}}\\
\resizebox{8.0cm}{!}{\includegraphics{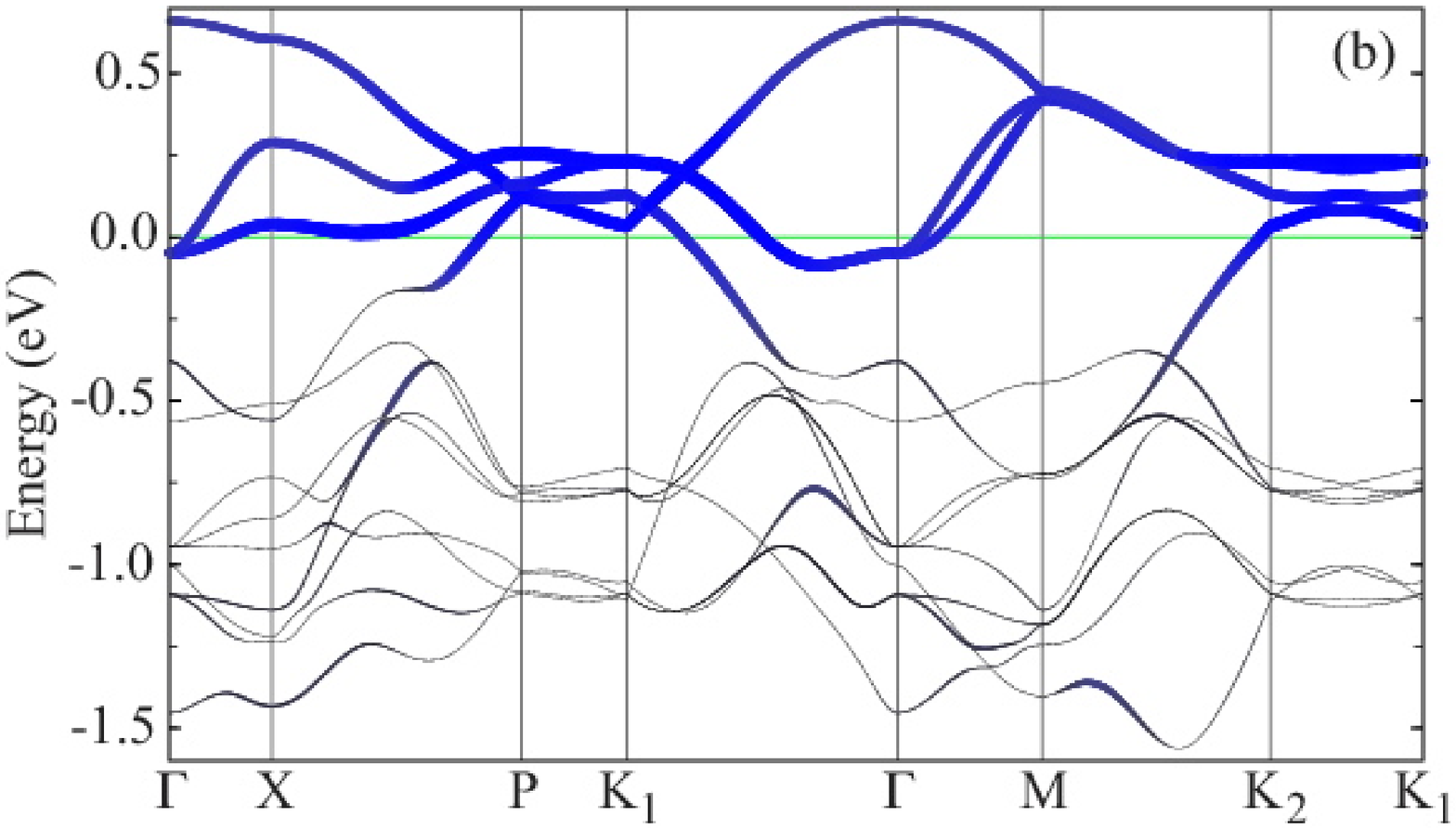}}\\
\caption{(Color online) Calculated majority-spin band 
dispersion of K$_2$Cr$_8$O$_{16}$ near the Fermi level.  
There are 12 $t_{2g}$ bands of the Cr $3d$ orbitals, 
3 of which cross the Fermi level.  
The line width is in proportion to the weight of the 
$d_{xy}$ component of Cr in (a) and to that of the 
$2p_z$ component of O(2) in (b).}
\label{fig5}
\end{center}
\end{figure}

\subsection{Fermi surface}

The calculated Fermi surfaces of K$_2$Cr$_8$O$_{16}$ in 
the ferromagnetic state are shown in Fig.~6.  
There are 12 $t_{2g}$ bands, 3 of which cross the Fermi 
level and form the semimetallic Fermi surfaces; i.e., 
the second and third bands (counted from the top) form 
the electron Fermi surfaces (see Figs.~6 (b) and (c)) and 
the fourth band forms the hole Fermi surface (see Fig.~6(a)).  
The wave functions at the Fermi surfaces have predominantly 
$d_{yz}$ and $d_{zx}$ character with large admixture 
of the O(2) $2p_z$ states as shown above.  

We find in Fig.~6(a) that there is a pair of the 1D-like 
parallel Fermi surfaces, which are seen to have a very 
good nesting feature.  The nesting vector is aligned roughly 
along the $\Gamma$-K$_1$ direction and has the value 
${\bm q}^*\simeq(0,0,0.147)2\pi/c$ or $(0,0,0.853)2\pi/c$ 
(a deviation in the $(q_x,q_y)$ component will be discussed 
below).  
Thus, the Fermi-surface instability corresponding to 
the wavenumber ${\bm q}^*$, leading to formation of the 
incommensurate, long-wavelength (with a period of $\sim$7c 
in the real space) DW, may be relevant with the opening of 
the charge gap in the present material.  
Note that the spin and charge DWs occur simultaneously 
with the same wavenumber ${\bm q}^*$ since we have only 
the up-spin electrons.  

\begin{figure}[tbh]
\begin{center}
\resizebox{7.9cm}{!}{\includegraphics{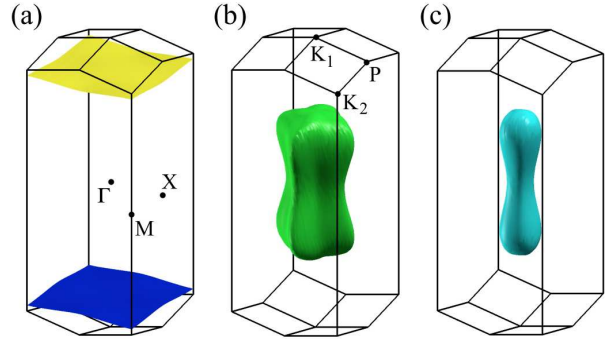}}\\
\caption{(Color online) Calculated Fermi surfaces of 
K$_2$Cr$_8$O$_{16}$ in the ferromagnetic state.  
The 61st to 63rd bands counted from the lowest are 
shown in (a) to (c), respectively.}
\label{fig6}
\end{center}
\end{figure}

\subsection{Generalized susceptibility}

To confirm the nesting features more precisely, we 
calculate the generalized susceptibility defined as 
$\chi_0({\bm q})=
\sum_{\bm k}\big(f(E_{\bm k})-f(E_{\bm k+q})\big)
/(E_{\bm k+q}-E_{\bm k})$,\cite{rath} where $E_{\bm k}$ 
is the band dispersion and $f$ is the Fermi distribution 
function at $T=0$ K.  The calculated result is shown 
in Fig.~7, where the contribution from only the 61st 
band is given.  Contributions from other bands including 
interband contributions are rather monotone functions 
of ${\bm k}$ over the entire Brillouin zone.  
Thus, the peak structure coming from the 61st band 
remains as a maximum even in the total susceptibility 
estimated in the constant matrix-element approximation.  

To be more precise, the peak structure at 
$q_z^*=0.295\pi/c$ and $1.705\pi/c$ seen in Fig.~7 
remains strong, irrespective of the value of $(q_x,q_y)$, 
although there is a small variation in the 
$(q_x,q_y)$ plane.  The true maximum appears 
at ${\bm q}^*\simeq(\pi/a,\pi/a,q_z^*)$, or around 
$(\pi/a,\pi/a,q_z^*)$, splitting and deviating slightly 
from $(\pi/a,\pi/a,q_z^*)$ in the $(q_x,q_y)$ plane.  
Thus, if we include the effects of electron 
correlations, the susceptibility can diverge at 
this momentum ${\bm q}^*$, resulting in the 
formation of the incommensurate, long-wavelength 
charge and spin DW, which we hope will be checked 
by experiment in near future.  

\begin{figure}[tbh]
\begin{center}
\resizebox{7.4cm}{!}{\includegraphics{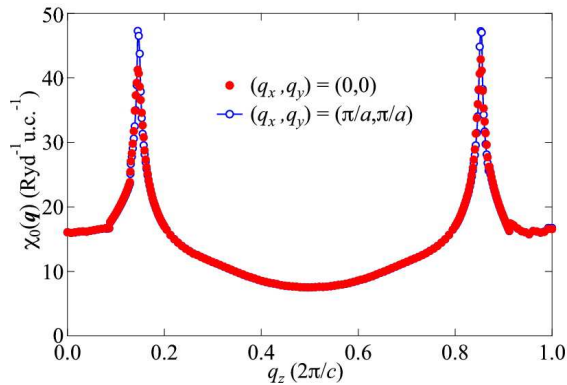}}\\
\caption{(Color online) Generalized susceptibility 
$\chi_0({\bm q})$ for the noninteracting band structure 
of K$_2$Cr$_8$O$_{16}$.  Contribution from only the 61st 
band is shown.}
\label{fig7}
\end{center}
\end{figure}

\subsection{Metal-insulator transition}

Now, let us discuss possible mechanisms of the MIT observed 
in K$_2$Cr$_8$O$_{16}$ at $T_c\simeq 90$ K.  
There are two types of scenarios: (i) correlation scenario 
where the electron correlations play an essential role and 
(ii) band scenario where the band structure solely determines 
whether the system is metallic or insulating.  
As a correlation scenario, we may have a possibility of the 
formation of the incommensurate, long-wavelength charge and 
spin DW due to the Fermi-surface nesting as discussed above.  
This may give rise to the opening of the charge gap, 
resulting in the MIT observed in the present material.  
As another correlation scenario, we may have the charge ordering 
(CO) if the intersite Coulombic repulsions between charge carriers 
are sufficiently strong.  In the present system, there may 
occur the localization of 3 spinless $t_{2g}$ holes per 4 Cr 
site.  We find that the possible spatial CO patterns are 
equivalent to those of 
K$_2$V$_8$O$_{16}$,\cite{isobe,horiuchi,sakamaki} which results 
in the doubling of the unit cell along the $c$ axis.  
As a band scenario, we may have the semimetal-to-semiconductor 
transition in the present system with the integer filling 
of spinless fermions as discussed in Sec.~III C.  Here, no 
spatial localization of carriers occurs, but the band gap may 
open if the 4-fold rotational symmetry around the $c$ axis is 
broken due to lattice distortions and the distortion is 
sufficiently large.  

Although the lattice distortion is essential for the MIT in 
any of the above cases and its identification can specify 
which scenario is realized, no structural distortions 
associated with the metal-insulator transition have been 
observed so far.\cite{hasegawa}  This seems to suggest that 
the simple distortions such as the doubling of the unit cell 
along the $c$ axis and the breaking of the 4-fold rotational 
symmetry around the $c$ axis may be ruled out.  
We therefore argue that the formation of the incommensurate, 
long-wavelength DW, the observation of which is sometimes 
not very easy, may be relevant with the MIT in the present 
material.  
We hope that the anomaly at ${\bm q}*$ estimated above 
will be sought for carefully in future experiment.  

Another experimental fact to be noted is that the spin 
polarization is hardly affected by the MIT in this material.  
This is interesting because the double-exchange mechanism for 
ferromagnetism (of which the energy gain occurs due to the 
first-order process of the hopping) should cease to work if 
the charge gap opens clearly in the majority-spin band and 
coherent motion of carriers vanishes.  The double-exchange 
ferromagnetism may be killed in such cases.  However, if 
the charge gap is small enough, the second-order processes 
of the hopping of conduction electrons can give rise to 
the ferromagnetic interaction between localized spins as well 
and thus the ferromagnetism can be maintained.  The vanishing 
first-order process seems to affect very little on the 
already fully spin-polarized system.  
Experimentally,\cite{hasegawa} the electrical resistivity 
measurement shows that the temperature dependence is of a 
three-dimensional variable-range-hopping type rather than a 
thermal activation type and thus it is not clear whether the 
well-defined charge gap is present in this material.  
Also, the charge gap expected in the formation of the 
incommensurate, long-wavelength DW state corresponding to 
${\bm q}^*$ may not be so large.  
The observed saturated ferromagnetism maintained under the 
gap formation seems to be related to such situations.  

\subsection{Negative charge-transfer gap}

We here point out that the present material K$_2$Cr$_8$O$_{16}$ 
is a doped negative charge-transfer-gap (CT-gap) type 
system\cite{khomskii} in the Zaanen-Sawatzky-Allen phase 
diagram.\cite{zaanen}  
Because K$_2$Cr$_8$O$_{16}$ is in the mixed-valent state, 
we need not invoke the negative CT-gap situation\cite{khomskii} 
for metallization of the system, as is realized in CrO$_2$ 
where the metallization via self-doping mechanism 
operates.\cite{korotin}  
However, in a hypothetical K$_0$Cr$_8$O$_{16}$ system, where 
K ions are completely depleted and all the Cr ions are in 
the $d^2$ state, this self-doping mechanism should be 
relevant for both metallization and double-exchange 
ferromagnetism unless the strong electron correlations leads 
the system to a Mott-insulating ground state with spin $S=1$ 
local moments aligned antiferromagnetically or with local 
spin-singlet pairs of $S=1$ spins.  
We here point out that, even in K$_2$Cr$_8$O$_{16}$, the 
negative CT-gap situation is realized between the O(2) $2p$ 
and Cr $3d$ $t_{2g}$ orbitals, as is evident in the calculated 
negative spin-polarization of the O(2) ions.  
In fact, results of our GGA+$U$ calculation for 
K$_2$Cr$_8$O$_{16}$ indicate that the gap in the DOS 
opens when $U$ is large enough although the Fermi level 
is away from the gap due to the mixed-valent nature of the 
system.  
Thus, the deficiency of K ions, if realized experimentally, 
can offer an important opportunity for studying the 
negative CT-gap situation and its carrier number dependence.  

%%%%%%%%%%%%%%%%%%%%%%%%%%%%%%%%
\section{Summary and prospects}
%%%%%%%%%%%%%%%%%%%%%%%%%%%%%%%%

In summary, we have made the first-principles electronic 
structure calculations and predicted that a chromium oxide 
K$_2$Cr$_8$O$_{16}$ of hollandite type should be a half-metallic 
ferromagnet.  We have shown that the double-exchange 
mechanism is responsible for the observed saturated 
ferromagnetism.  We have discussed the possible scenarios 
of the metal-insulator transition observed at low 
temperature, which include possibilities of the formation 
of the charge order and semimetal-to-semiconductor 
transition, but we have argued that the formation of the 
incommensurate, long-wavelength density wave of spinless 
fermions caused by the Fermi-surface nesting may be the 
origin of the opening of the charge gap.  
We hope that these predictions will be checked by further 
experimental studies.  

Before closing this paper, let us discuss some prospects 
for future studies of this material, which may include the 
following:  
\\
(i) This material offers an interesting opportunity for 
studying the effects of the majority-spin band gap on 
the electronic properties of half-metallic ferromagnets.  
In particular, the so-called non-quasiparticle states 
\cite{edwards,irkhin} and the effects of the opening of 
the gap on the non-quasiparticle states should be pursued 
by the spectroscopic experiments such as photoemission 
and X-ray absorption 
experiment.\cite{tsujioka,huang1,huang2,kurmaev}
\\
(ii) The transport properties of K$_2$Cr$_8$O$_{16}$ are also 
anomalous,\cite{hasegawa} as those of a 
``bad metal'',\cite{ranno,irkhin,suzuki,mazin,watts} which 
may also be interesting because not only the spin fluctuations 
but also the collective fluctuations of the orbital degrees 
of freedom between the degenerate $d_{yz}$ and $d_{zx}$ 
orbitals may play an important role.  
\\
(iii) Introduction of the deficiency of K ions, if it could 
be made experimentally, would enable us to examine the 
effects of changing the doping rate on the gap formation, 
double exchange ferromagnetism, and half metallicity.  
We also point out that only the 1D-like Fermi surface 
should remain by $\sim$10\% hole doping if we assume 
the rigid-band shift of the Fermi level (see Fig.~5), 
suggesting the realization of the 1D electron system.  
\\
(iv) Theoretically, one should study the three-band Hubbard 
model with the $d_{xy}$, $d_{yz}$, and $d_{zx}$ orbitals to 
see, in particular, the effects of orbital fluctuations on the 
electronic states of the system.  The study may be extended 
to the $d$-$p$ model including the $2p$ orbitals of O ions 
if one wants to take into account the negative CT-gap situation 
of the present system.  Techniques beyond the GGA+$U$ method, 
such as the dynamical mean-field theory\cite{craco} and 
variational cluster approach,\cite{chioncel} are required 
to examine the presence of non-quasiparticle states and 
their deformation due to the gap formation.  

Finally, we want to point out that, because the basic electronic 
states of the present system have many aspects in common with 
those of CrO$_2$, it may be very useful to clarify similarities 
and differences in the electronic states between 
K$_2$Cr$_8$O$_{16}$ and CrO$_2$ in all the above respects.  

\begin{acknowledgments}
We would like to thank M. Isobe, T. Yamauchi, and Y. Ueda 
for informative discussions on experimental aspects of 
K$_2$Cr$_8$O$_{16}$ and S. Ishihara and T. Shirakawa for useful 
discussions on theoretical aspects.  
This work was supported in part by a Grant-in-Aid for 
Scientific Research (No.~19014004) from the Ministry 
of Education, Culture, Sports, Science and Technology 
of Japan.  
A part of computations was carried out at the 
Research Center for Computational Science, 
Okazaki Research Facilities, and the Institute 
for Solid State Physics, University of Tokyo.  
\end{acknowledgments}

\end{document}